\documentclass[twocolumn,showpacs,preprintnumbers,amsmath,amssymb]{revtex4}

\usepackage{graphicx}
\usepackage{dcolumn}
\usepackage{bm}
\begin{document} 

\title{ARPES spectra of the stripe phase in the 2D t-J model}
\author{R. Eder$^1$ and Y. Ohta$^2$}
\affiliation{$^1$Institut f\"ur Festk\"orperphysik, Forschungszentrum
  Karlsruhe, 76021 Karlsruhe, Germany\\
$^2$Department of Physics, Chiba University, Chiba 263-8522, Japan}
\date{\today}

\begin{abstract}
The 2D t-J model with and without $t'$ and $t''$ hopping-terms is
studied by exact diagonalization on a $5\times 4$ cluster, which
realizes a hole stripe in $y$-direction in a spin-Peierls.
Next nearest hopping terms with a sign appropriate 
for hole-doped cuprates enhance the stripe formation. 
The dispersion of the quasiparticle-peaks in the single-particle 
spectrum is in good agreement with bond operator theory for hole
motion in the spin-Peierls phase, particularly so for realistic 
values of $t'$ and $t''$. The resulting spectral weight distribution and 
Fermi surface agree well with experimental ARPES spectra on 
La$_{1.28}$Nd$_{0.6}$Sr$_{0.12}$CuO$_4$.
\end{abstract} 

\maketitle
\begin{figure}
\includegraphics{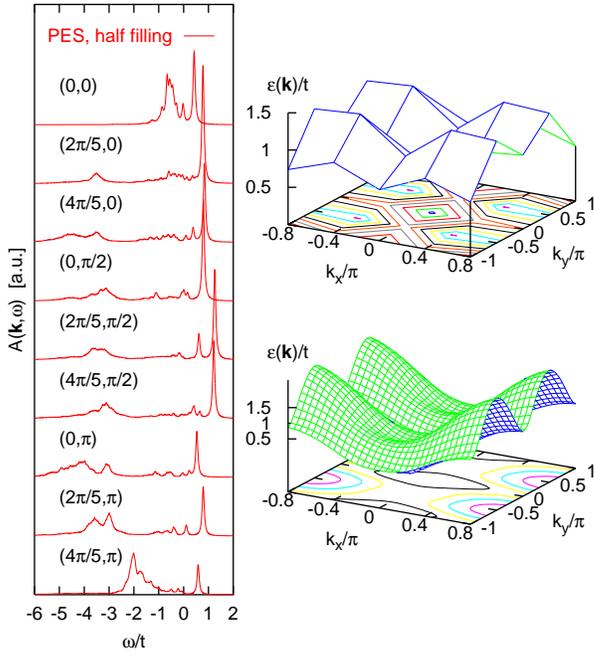}
\caption{\label{fig1} Left: Photoemission spectrum (PES) for the half-filled
$5\times 4$ cluster, $J/t=0.5.$, $t'/t=-0.2$, $t''/t=0.1$.\\
Right: Dispersion of the quasiparticle peak 
as extracted from the numerical spectra (top) compared to
the theoretical single hole dispersion $\epsilon_1({\bf k})$
in the spin-Peierls phase (bottom).
}
\end{figure}
\begin{figure}
\includegraphics{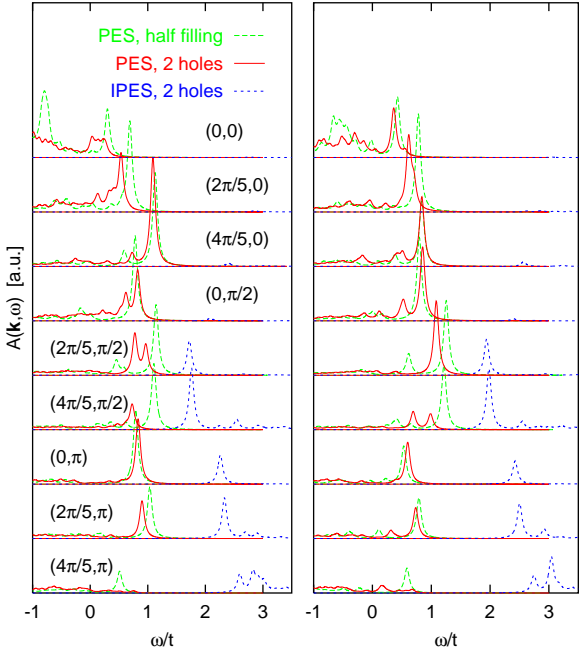}
\caption{\label{fig2} Single particle spectrum for the 
$5\times 4$ cluster with two holes.
Parameter values are  $J/t=0.5.$, $t'=t''=0$ (Left)
and  $J/t=0.5.$, $t'/t=-0.2$, $t''/t=0.1$ (Right).
Only the parts near the Fermi energy $E_F/t\approx 1.5$ are shown.}
\end{figure}

The prediction of charged stripes\cite{mf} and their subsequent
experimental verification\cite{Tranquada} stand out as one of the 
rare instances,
where a nontrivial theoretical prediction for cuprate superconductors
was found consistent with experiment. Accordingly, there is currently
considerable interest in the mechanism leading to the formation
and the physical implications of stripes
\cite{Loew,TsunetsuguTroyerRice,Zaanen,WhiteScalapino,VojtaSachdev,Chernyshev,Martinsetal,CastroNeto}.
On the other hand few experimental techniques provide such direct
experimental insight into the electronic structure of a given compound
as angle resolved photoemission spectroscopy
(ARPES). It is therefore quite natural to look for the fingerprints
of stripes in ARPES spectra and indeed the results of Zhou {\em et al.}
on La$_{1.28}$Nd$_{0.6}$Sr$_{0.12}$CuO$_4$\cite{Zhou} are widely 
considered as strong evidence for
stripes. It is the purpose of the present manuscript to
present single particle spectra obtained by `computer spectroscopy'
on the stripe phase of the 2D t-J model, presumably the simplest theoretical
description of the CuO$_2$ planes in cuprate superconductors.
As will be seen below these result combined with a relatively crude
theory for hole motion in a spin-Peierls `background' already give a quite
satisfactory description of most of the experimental results.\\
The t-J model reads
\[
H = - \sum_{i,j}\sum_{\sigma} t_{i,j}
\hat{c}_{i,\sigma}^\dagger \hat{c}_{j,\sigma} 
+ J\sum_{\langle i,j\rangle} 
\left(\vec{S}_i \cdot \vec{S}_{j} - \frac{n_i n_j}{4}\right)
\]
Thereby $\langle i,j\rangle$ denotes summation over pairs of
nearest neighbor sites,
$\hat{c}_{i,\sigma} = c_{i,\sigma} (1-n_{i,\bar{\sigma}})$ and
$\vec{S}_i$ and $n_i$ denote the operators of electron spin and
density at site $i$, respectively. We denote the hopping matrix elements
$t_{i,j}$ between $(1,0)$-like neighbors by $t$, between
$(1,1)$-like neighbors by $t'$ and between $(2,0)$-like neighbors
$t''$, all other $t_{i,j}$ are zero. Throughout we will assume that
$t'/t''=-2$, as would be appropriate if the physical origin of these
terms is hopping via the apex oxgen $2p_zs$ orbital\cite{Raimondi}, and
$t'/t<0$, as is the case for hole-doped compounds.\\
The method we apply to study this model is exact diagonalization
of finite clusters by means of the Lanczos algorithm\cite{Dagotto}.
In a preceding paper\cite{5x4} we have shown that by changing the
geometry of the cluster from the standard tilted square form
to a rectangular one (more precisley: to a $5\times 4$ cluster)
a ground state with a pronounced stripe like charge inhomogeneity emerges.
Here we want to discuss the single particle spectra of this state.\\
A question to worry about first is, whether the stripes survive the
additional hopping terms $\propto t',t''$. Intuitively 
this is not what one would
expect, because additional hopping terms increase the mobility of the holes,
whence any spatial inhomogeneity should be washed out more efficiently.
Surprisingly enough, the numerics show, 
that exactly the opposite is happening: the additional hopping terms
even slightly enhance the charge inhomogeneity. This is demonstrated
in Table \ref{tab1}, which compares the static density correlation
function $g_D({\bf R}) = \sum_j \langle n_{j} n_{j+{\bf R}} \rangle$
for vanishing and finite $t'$ and $t''$.
Next nearest-neighbor hopping terms with the proper sign for
hole-doped cuprates thus seem to have a stabilizing effect
on stripes - if any. \\
\begin{table}[b]
  \begin{center}
\begin{tabular}{|r r|rrr|rrr|}
\hline
              &  2 & 0.298  & 0.139 &  0.035 &  0.303 &  0.151 &  0.037 \\
$R_y\uparrow$ &  1 & 0.239  & 0.144 &  0.041 &  0.261 &  0.144 &  0.043 \\
              &  0 & 2.000  & 0.049 &  0.018 &  2.000 &  0.016 &  0.010 \\
\hline
              &    &  0     &  1    &   2    &   0    &   1    &   2    \\
&    &        &  $R_x \rightarrow$ & &       & $R_x \rightarrow$ & \\
\hline    	     	           	          	   
\end{tabular}
\caption{Static density correlation function $g_D({\bf R})$, 
$5\times 4$ cluster with $2$ holes, $J/t=0.5$.
Other parameters are $t'=t''=0$ (left panel) and
$t'/t=-0.4$, $t''/t=0.2$ (right panel).}
\label{tab1}
\end{center}
\end{table}
Next we adress a special feature of the $5\times 4$ cluster,
which will be essential to understand the hole dynamics,
namely the presence of spin-Peierls dimerization
even at half filling. Table \ref{tab2} shows that the static spin
correlation function is strongly anisotropic, with
singlet-bonds predominantly in $y$-direction. 
Since the boundary conditions in
$5\times 4$ frustrate the N\'eel order, they apparently stabilize the
energetically close spin-Peierls phase. In fact, the GS energy of the 
$5\times 4$ cluster is only marginally higher than that of the 
square-shaped  $\sqrt{20}\times  \sqrt{20}$ cluster 
($-1.165 J/$site  vs.  $-1.191 J/$site).
Clearly, this is a confirmation of the proposal by
Read and Sachdev\cite{ReadSachdevI} that the transition to a spin-Peierls
phase is a likely instability of the $S=\frac{1}{2}$
2D Heisenberg antiferromagnet.\\
\begin{table}[b]
  \begin{center}
\begin{tabular}{|r r|rrr|}
\hline
              &  2 & 0.222  &-0.162 &  0.052 \\
$R_y\uparrow$ &  1 &-0.389  & 0.173 & -0.059 \\
              &  0 & 0.750  &-0.276 &  0.061 \\
\hline
              &    &  0     &  1    &   2    \\
&    &        &  $R_x \rightarrow$ &         \\
\hline    	     	           	          	   
\end{tabular}
\caption{Static spin correlation function for the half-filled $5\times 4$
cluster.}
\label{tab2}
\end{center}
\end{table}
Clear evidence for the spin-Peierls nature of the half-filled
ground state can be seen in the single particle spectral function
$A({\bf k},\omega)$ (see Ref. \cite{Dagotto} for a definiton),
which is shown in Figure \ref{fig1}.
It is immediately obvious that this differs markedly
from the familiar dispersion for a hole in an antiferromagnet:
whereas for hole motion in a N\'eel state the top of
the ARPES spectrum is at 
$(\frac{\pi}{2},\frac{\pi}{2})$\cite{Wells,t_prime_refs},
the dispersion seen in the half-filled
$5\times 4$ cluster has its maximum at $(\frac{4\pi}{5},\pi/2)$
- which probably means $(\pi,\pi/2)$ in the infinite system.
Another notable feature is the symmetry of the dispersion
under the exchange
$(k_x,0) \rightarrow (k_x,\pi)$ - which is exactly what one would expect 
from the doubling of the unit cell by spin-Peierls order with
dimers in $y$-direction. 
To be more quantitative, let us discuss
the single-hole dispersion in the spin-Peierls phase. 
Starting from a product state of
singlets, which cover the bonds of the lattice in the form
of a columnar pattern, we derive a Hamiltonian for
the motion of singly occupied dimers. The singlet state on
a bond formed by the sites $(1,2)$ is
$|s\rangle = \frac{1}{\sqrt{2}} (
\hat{c}_{1,\uparrow}^\dagger \hat{c}_{2,\downarrow}^\dagger -
\hat{c}_{1,\downarrow}^\dagger \hat{c}_{2,\uparrow}^\dagger )|0\rangle$ .
A dimer with a single hole can be in either the bonding or
antibonding state:
$|\pm,\sigma\rangle = \frac{sign(\sigma)}{\sqrt{2}} (
\hat{c}_{1,\sigma}^\dagger \pm \hat{c}_{2,\sigma}^\dagger  )|0\rangle$.
Introducing the `creation operator'
$h_{\pm,\sigma}^\dagger = |\pm,\sigma\rangle \langle s|$
we can - by straightforward generalization
of Refs. \cite{ladder,Sushkov_doped,Park_Sachdev} - write down the
following Hamiltonian
describing the motion of these effective Fermions:
\begin{eqnarray}
H &=& \sum_{{\bf k},\sigma}
\epsilon_+({\bf k})\; h_{+,{\bf k},\sigma}^\dagger  h_{+,{\bf k},\sigma}^{\;} +
\epsilon_-({\bf k})\; h_{-,{\bf k},\sigma}^\dagger  h_{-,{\bf k},\sigma}^{\;}
\nonumber \\
&& \;\; \; \; \; \; \; \;  + \left(V({\bf k})\; h_{+,{\bf k},\sigma}^\dagger 
h_{-,{\bf k},\sigma}^{\;} + H.c.\right)\nonumber \\
\epsilon_\pm({\bf k}) &=& \mp t
+ t\left(\cos(k_x)\pm\frac{\cos(2k_y)}{2}\right)
\nonumber \\
&& \;\; \; \; \; \; \; \; \pm t'\cos(k_x)\left(1+\cos(2k_y\right)) \nonumber \\
&&  \;\; \; \; \; \; \; \;\;\; \; \; \; \; \; \; 
+ t''\left(\cos(2k_x) +\cos(2k_y)\right)\nonumber \\
V({\bf k}) &=& -\frac{it}{2} \sin(2k_y) -it' \cos(k_x)\sin(2k_y).
\label{ham}
\end{eqnarray}
Here the coupling of the dimer-Fermions
to triplet-excitations of the
bonds\cite{ladder,Sushkov_doped,Park_Sachdev} has been
neglected for simplicity.
Diagonalizing the Hamiltonian  (\ref{ham}) we obtain the dispersion 
relation $\epsilon_\alpha({\bf k})$ and the quasiparticles
$\gamma_{\alpha,{\bf k},\sigma}^\dagger
= u_{\alpha,{\bf k}} h_{+,{\bf q},\sigma}^\dagger
+  v_{\alpha,{\bf k}} h_{-,{\bf q},\sigma}^\dagger$,
where $\alpha \in \{\/1,2\}$.
Figure \ref{fig1} shows that there is good agreement
between the numerical peak dispersion in the $5\times 4$ cluster 
and our simple theory. The main differences are the  flattening
of the cluster dispersion near the band maximum at $(\frac{4\pi}{5},\pi/2)$
and the smaller bandwidth
in the numerical spectra, which is probably due to the coupling
of the Fermions to triplet 
excitations\cite{ladder,Sushkov_doped,Park_Sachdev}.
Taking into account the simplicity of our calculation, however,
the agreement is quite satisfactory.
Summarizing our discussion so far we have seen that the
$5\times 4$ cluster shows 
a single-hole dispersion that differs markedly from that for
a N\'eel background, but is in good agreement with a simple
bond-operator calculation for hole-motion in a spin-Peierls phase.
This reflects the fact that
the frustration of N\`eel order along the odd-numbered side of the
$5\times 4$ has driven the system into the spin-Peierls phase of the
2D Heisenberg antiferromagnet, consistent with
arguments given by Read and Sachdev\cite{ReadSachdevI}.\\
While the dimerization clearly renders the $5\times 4$ cluster
unsuitable to describe undoped compounds
such as Sr$_2$CuO$_2$Cl$_2$\cite{Wells}, it 
makes the interpretation of the spectra for the striped ground state
at finite doping a lot easier - as will be seen now.
Ignoring for the moment the 
formation of a hole stripe as well as the fact that the
dimer-Fermions $h_{\pm,{\bf q},\sigma}^\dagger$
actually obey a hard-core constraint,
one would expect, that the doped holes accumulate
near the top of the single-hole dispersion, thus forming to simplest
approximation a cigar-shaped hole pocket\cite{Sushkov_doped,Park_Sachdev}
centered at $(\frac{4\pi}{5},\frac{\pi}{2})$ 
(or rather $(\pi,\frac{\pi}{2})$ in the thermodynamical limit).
Figure \ref{fig2}, which shows the single particle spectrum for the
two-hole GS for different $t'$ and $t''$, demonstrates that this is 
indeed exactly what happens.
There is a clear analogy between the spectra for the doped
case and half-filling, the dispersion of the quasiparticle peak
being essentially unchanged. There are two major changes, both of
whom could have been expected: first, peaks with a higher binding
energy become rather diffuse, which is nothing but the familiar
Landau-damping. Second, the peaks at $(\frac{4\pi}{5},\frac{\pi}{2})$ and 
(to a lesser extent) at
$(\frac{2\pi}{5},\frac{\pi}{2})$ cross to the IPES spectrum.
It should be noted that due to finite-size effects there is always
a significant gap between the PES and IPES spectrum of finite
clusters - after all the electron numbers of
initial and final state differ by a finite fraction 
(10$\%$ in the present case). It is therefore
impossible to decide, whether the gap is simply a finite-size effect
or due to the stripe formation.
The only deviation from this ideal rigid-band behaviour is the
appearance of high-energy IPES peaks
along $(0,\pi)\rightarrow (\frac{4\pi}{5},\pi)$. The interpretation of these
peaks, however, is straightforward: in inverse photoemission, an
electron is necessarily inserted into a dimer occupied by a single
electron. The spins of the two electrons then can couple either to a singlet
- which means the IPES process leads back to the spin-Peierls `vacuum' -
or to a triplet - which means the IPES process leaves the system in a 
spin-excited state. The IPES peaks along $(0,\pi)\rightarrow
(\pi,\pi)$ presumably originate from the latter process
(we note that exactly the same holds true also for the `usual'
ground state of the t-J model, see \cite{IPES}).
Taken together, the data presented so far
demonstrate, that the spin-Peierls order in the striped phase is the
key to understand its single particle spectra.
Neglecting subtleties such as the possible formation of a
Luttinger liquid along the stripes\cite{Zaanen}, the possible condensation
of $d$-like hole pairs along the
stripes\cite{WhiteScalapino_unpublished} 
or the formation of various kinds of order
parameter\cite{Park_Sachdev}, the spectra show, that
the system can be described by the dispersion for a single hole
in a spin-Peierls background being filled up with holes.
Thereby the momentum $(\pi,\frac{\pi}{2})$
where the band crosses the Fermi energy is
independent of $t'$ and $t''$ -
the reason is simply that irrespective of $t'$ and $t''$
the dispersion is symmetric under reflection by
the line $(0,\frac{\pi}{2}) \rightarrow (\pi,\frac{\pi}{2})$
due to the underlying spin-Peierls order.\\
\begin{figure}
\includegraphics{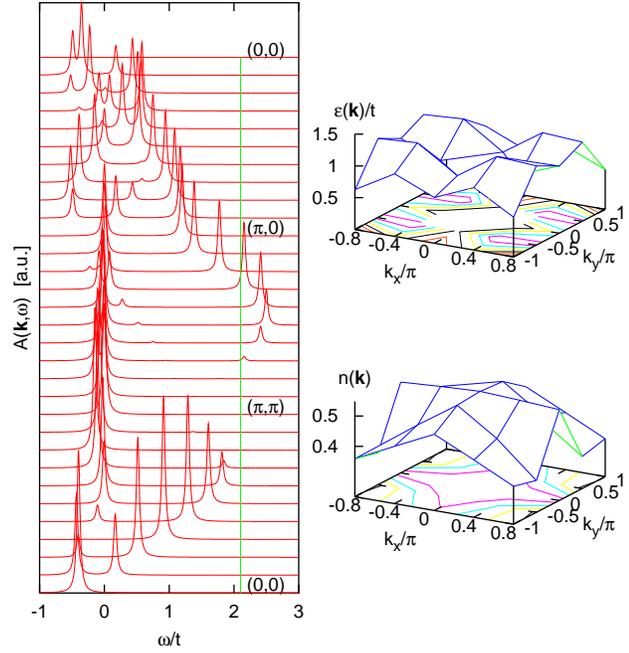}
\caption{\label{fig3} Left: Single-particle spectral function from
bond-operator theory for $t'/t=-0.1$, $t''/t=0.05$,
the Fermi energy (verical line) corresponds to a hole density of $0.16$.\\
Right: Single hole dispersion (top) and momentum distribution
$n({\bf k})$ (bottom) for $5\times 4$ and the same
parameters.}
\end{figure}
We now want to use this (rather oversimplified)
scenario to discuss the experimental ARPES spectra
on La$_{1.28}$Nd$_{0.6}$Sr$_{0.12}$CuO$_4$\cite{Zhou}.
To compute the full ARPES spectrum 
we need the following representation of the
electron annihilation operator:
\begin{equation}
c_{{\bf k},\sigma}= \frac{1}{\sqrt{2}}\left(\cos(\frac{k_y}{2})\; h_{+,-{\bf q},\bar{\sigma}}^\dagger
- i\sin(\frac{k_y}{2})\; h_{-,-{\bf q},\bar{\sigma}}^\dagger\right).
\label{weight}
\end{equation}
Thereby ${\bf q}$ is the `backfolded version' of ${\bf k}$,
so as to take into account that the Brillouin zone of the 
spin-Peierls phase is $[-\pi,\pi]\times[-\frac{\pi}{2},\frac{\pi}{2}]$.
Equation (\ref{weight}) is readily
verified by taking matrix elments of both sides
between $|s\rangle$ and $|\pm,\sigma\rangle$. The
spectral weight of a given quasiparticle branch then is 
$w=| u_{\alpha,{\bf k}}^*\;\cos(\frac{k_y}{2}) + i
v_{\alpha,{\bf k}}^*\;\sin(\frac{k_y}{2})|^2$.
In Figure \ref{fig3} the spectra for the
symmetric momenta $(k_x,k_y)$ and $(k_y,k_x)$ have been averaged,
as would be appropriate for an ARPES experiment on a compound with
domains of different singlet direction\cite{Zhou}.
The experimental ARPES spectrum should be compared to the parts
of the spectrum below the Fermi energy, indicated as the
vertical line in Figure  \ref{fig3}.
Along the $(1,1)$ direction there is a band dispersing upwards
and disappearing halfway between $(0,0)$ and $(\pi,\pi)$.
This is actually not a Fermi level crossing - the {\em dispersion}
of the band actually bends downwards again after passing
through $(\frac{\pi}{2},\frac{\pi}{2})$ - but rather a
vanishing of the spectral weight 
due to a destructive interplay between the coefficients
$(u_{\alpha,{\bf k}}, v_{\alpha,{\bf k}})$ 
and the `form factors' $(\cos(\frac{k_y}{2}),i\sin(\frac{k_y}{2}))$
of the dimer fermions.
The situation along $(1,1)$ thus is similar to the
`remnant Fermi surface' in the half-filled compounds\cite{Ronning}.
Along $(\pi,0)\rightarrow (\pi,\pi)$ there are 
two `real' Fermi surface crossings, which are symmetric around
$(\pi,\frac{\pi}{2})$. Again due
to matrix element effects, the one between $(\pi,\frac{\pi}{2})$ and
$(\pi,\pi)$ has small spectral weight, which would
probably render it unobservable in an ARPES experiment.
`Spectroscopically' the system thus would look very much like having
a single Fermi surface sheet near $(\pi,0)$ running
roughly parallel to $(1,0)$ (see Figure \ref{fig3}) 
and disappearing as $(\frac{\pi}{2},\frac{\pi}{2})$ is approached.
Compared to experiment, the band crossing along $(\pi,0)\rightarrow
(\pi,\pi)$ is too far away from $(\pi,0)$ - it should be noted,
however, that in the actual stripe phase the hole density is
inhomogeneous, so that the hole density `within' the stripes
- which probably determines the area of the cigar -
is higher than average. Clearly, this would shift the Fermi surface
crossing towards $(\pi,0)$.
No crossing should be seen along $(1,1)$, only a band dispersing
upwards and disappearing somehwere below $E_F$.
The agreement with experiment would be almost perfect
if the band portion near $(\pi,0)$ were slightly moved upward
compared to the band maximum near  $(\frac{\pi}{2},\frac{\pi}{2})$ -
such fine details are probably beyond our very simple
bond-operator calculation.
To compare to the data of Zhou {\em et al.} Figure
\ref{fig3} also shows the momentum distribution
$n({\bf k})= \int_{-\infty}^\mu A_-({\bf k},\omega)d\omega$
obtained from the cluster diagonalization. This shows some
anisotropy, with in particular a `ridge' of almost constant
$n({\bf k})$ running along $(1,0)$. The cross-shaped
character is not as pronounced as in the experiment,
but this is simply due to the `partial occupation'
of the momentum $(\frac{2\pi}{5},\frac{\pi}{2})$, which
crosses only partially from PES to IPES, see Figure \ref{fig2},
nd thus retains a relatively large $n({\bf k})$.
Nevertheless one may say, that when combined with the
simple bond-operator theory, the numerical data give
a good description of the experimental
data on  La$_{1.28}$Nd$_{0.6}$Sr$_{0.12}$CuO$_4$.\\
In summary, numerically exact diagonalization results show,
that the 2D t-J model with next-nearest neighbor hopping
terms appropriate to describe hole doped cuprates
has a spin-Peierls phase which leads to the formation of pronounced
hole-stripes. Given that stripes are an experimental fact
in cuprates\cite{Tranquada} and that the 
density matrix renormalization group calculations on much 
larger clusters of the t-J model\cite{WhiteScalapino,Martinsetal} 
have also shown clear
evidence for stripes, one may be confident that these hole-stripes are
indeed representative for those in the bulk system.
The single particle spectrum in the stripe phase then is found to be in
good agreement with a simple bond-operator theory for hole motion
in a spin-Peierls phase, thus providing further evidence
for the intimate relationship between spin-Peierls odering and
stripe formation.
Upon doping, holes accumulate near the top of the
single-hole dispersion, to simplest approximation forming
cigar-shaped pockets centered
on the corner of the spin-Peierls Brillouin zone at
$(\pi,\frac{\pi}{2})$ - the latter in full agreement with
bond-operator theory\cite{Sushkov_doped,Park_Sachdev}.
The notion of a Fermi
surface should not be taken too literal,
because close to $E_F$ the stripe formation is likely to change
this simply free-particle picture drastically, but
all in all the quasiparticle dispersion and
spectral weight distribution of the stripe phase
as seen in the simulations
are in good agreement with ARPES experiments.

\end{document}